\documentclass[seceq,preprint]{ptptex}
\usepackage{amsmath,amssymb,latexsym}
\usepackage[dvips]{graphicx}

\newcommand{\mkk}{M_{\rm KK}}
\newcommand{\ket}{\left|p\right>}
\newcommand{\AD}[1]{$\ol{\mbox{D~\,}}\!\!\!$#1}

\newcommand{\ol}{\overline}
\newcommand{\ul}{\underline}
\newcommand{\ra}{\rightarrow}

\newcommand{\bR}{\mathbb R}
\newcommand{\bZ}{\mathbb Z}

\newcommand{\cC}{\mathcal C}
\newcommand{\cO}{\mathcal O}

\newcommand{\wt}{\widetilde}
\newcommand{\del}{\partial}
\newcommand{\nn}{\nonumber}

\def\drawbox#1#2{\hrule height#2pt
        \hbox{\vrule width#2pt height#1pt \kern#1pt
              \vrule width#2pt}
              \hrule height#2pt}

\def\Fund#1#2{\vcenter{\vbox{\drawbox{#1}{#2}}}}
\def\Asymm#1#2{\vcenter{\vbox{\drawbox{#1}{#2}
              \kern-#2pt       
              \drawbox{#1}{#2}}}}
\def\TAsymm#1#2{\vcenter{\vbox{\drawbox{#1}{#2}
              \kern-#2pt       
              \drawbox{#1}{#2}
              \kern-#2pt       
              \drawbox{#1}{#2}}}}

\def\Upbox#1#2{\vcenter{\vbox{\drawbox{#1}{#2}
              \kern+#2pt       
              \drawbox{#1}{0}}}}
\def\TUpbox#1#2{\vcenter{\vbox{\drawbox{#1}{#2}
              \kern+#2pt       
              \drawbox{#1}{0}
              \kern+#2pt       
              \drawbox{#1}{0}}}}

\def\fnd{\Fund{5.5}{0.4}}
\def\asym{\Asymm{5.5}{0.4}}
\def\Tasym{\TAsymm{5.5}{0.4}}
\def\sym{\fnd\kern-0.4pt\fnd}
\def\Tsym{\fnd\kern-0.4pt\fnd\kern-0.4pt\fnd}
\def\asymsym{\asym\kern-0.4pt\Upbox{5.5}{0.4}}
\def\Tasymsym{\Tasym\kern-0.4pt\TUpbox{5.5}{0.4}}

\notypesetlogo                       
\preprintnumber[3cm]{
IPMU 09-0085}

\title{
Mesons as Open Strings in a Holographic Dual of QCD
}

\author{
Toshiya \textsc{Imoto},$^1$\footnote{E-mail: 
{\tt imotoggl@gmail.com}}
Tadakatsu \textsc{Sakai},$^2$\footnote{E-mail: 
{\tt tsakai@eken.phys.nagoya-u.ac.jp}}
and Shigeki \textsc{Sugimoto},$^3$\footnote{E-mail: 
{\tt shigeki.sugimoto@ipmu.jp}}
}

\inst{
$^{1}$Daiwa Securities Capital Markets Co. Ltd.,\\
Grand Tokyo North Tower, Tokyo 100-6753, Japan\\
$^{2}$Department of Physics, Nagoya University,\\
 Nagoya 464-8602, Japan \\ 
$^{3}$Institute for the Physics and Mathematics of the Universe,\\
The University of Tokyo, Kashiwa 277-8568, Japan}

\abst{
We study meson spectrum obtained from massive open string modes in 
a holographic dual of QCD constructed 
on the basis of a D4/D8-brane configuration
in type IIA string theory. 
The spectrum includes mesons with spin higher than one.
Taking into account the effect of curved background perturbatively,
we obtain a mass formula for these mesons.
The mass formula exhibits the linear Regge
behavior at the leading order with subleading non-linear corrections.
We argue that the string spectrum
captures some features of the observed meson spectrum.
{}For example, $a_2(1320)$, $b_1(1235)$, $\pi(1300)$, $a_0(1450)$, etc.,
are identified as the first excited massive open string states and
$\rho_3(1690)$, $\pi_2(1670)$, etc.,  are identified as the second
excited states.
}

\begin{document}

\maketitle
\section{Introduction}

String theory was originally born as a theory of hadrons
in the late 1960s \cite{Nambu,Nie,Sus}.
One of the nice features of string theory
is that the spectrum exhibits the linear Regge behavior,
which is one of the mysterious properties found in the observed hadron
spectrum.
That is, the spin $J$ and mass squared $M^2$ of the hadrons lie on the
linear trajectory as
\begin{eqnarray}
J =\alpha_0+\alpha'M^2\ ,
\label{Regge}
\end{eqnarray}
where $\alpha_0$ and $\alpha'$ are parameters called
Regge intercept and Regge slope, respectively.
However, there were some difficulties in regarding
string theory as the fundamental theory of hadrons
at that time. For example, the space-time dimension of string theory
is higher than four, and the Regge intercepts turned out to be
inconsistent with the observed hadron spectrum.
In fact, the Regge intercepts for the leading trajectories are
$\alpha_0=1$ and $\alpha_0=2$ for open and closed strings,
respectively. This implies that there are massless particles with $J=1$
and $J=2$ in the open and closed string spectra, respectively.
The massless spin-two state in the closed string spectrum was later
interpreted as graviton \cite{Yoneya,schsch}, and string
theory has become a promising candidate of a unified theory
that includes quantum gravity.
Instead, QCD is now established as the fundamental theory of hadrons.

The situation has drastically
changed since the discovery of the gauge/string duality. 
(For a review, see Ref.~\citen{adscft}.)
It is conjectured that string theory in a certain ten-dimensional
curved space-time can be equivalent to a four-dimensional gauge theory.
Many examples of this type of duality have been found.
Since the space-time dimension in string theory description is
higher than four, it is often called a holographic description
of the four-dimensional gauge theory.
In some examples, it can be shown that massless particles
with $J=1,2$ in ten-dimensional string theory correspond to massive
particles in the four-dimensional gauge theory.
Therefore, the gauge/string duality has a potential to solve
the old problems in string theory as a theory of hadrons.

A holographic description of a gauge theory that realizes
four-dimensional $SU(N_c)$ QCD with $N_f$ massless flavors at low energy
was proposed in Ref.~\citen{SS1}. It is constructed from a D4/D8-brane
configuration in type IIA string theory.
The closed string sector is the same as that considered by Witten
in Ref.~\citen{Witten_D4}, in which a supergravity solution corresponding
to $N_c$ D4-branes wrapped on an $S^1$ with supersymmetry breaking
boundary condition is used to obtain a holographic description of
four-dimensional pure $SU(N_c)$ Yang-Mills theory.
This configuration has been used in the study of glueball
spectrum. (See, for example, Refs.~\citen{coot}-\citen{HaOz}).
As mentioned above, although the gravitons are massless in
the ten-dimensional curved space-time, they yield massive glueballs in
the four-dimensional world.
In addition, to incorporate $N_f$ flavors of quarks in the system,
$N_f$ D8-branes are placed in the Witten's D4-brane background as probes.
It was argued in Refs.~\citen{SS1} and \citen{SS2} 
that the open strings attached on the
D8-branes are interpreted as mesons and the low energy effective theory
is written as a five-dimensional $U(N_f)$ Yang-Mills - Chern-Simons (YM-CS)
theory. It turned out that this five-dimensional YM-CS theory
reproduces various phenomenological models constructed to describe
the properties of hadrons. Furthermore, many quantities such as masses
and couplings calculated in the holographic description
are roughly in good agreement with the experimental results.
This five-dimensional gauge field originates from the massless modes
in the open string spectrum. It produces a massless pseudo-scalar meson
(pion) together with an infinite tower of massive vector and
axial-vector mesons with $J^{PC}=1^{--}$ and $1^{++}$, where $P$
and $C$ are parity and charge conjugation parity, respectively.
Clearly, it can only cover a part of the whole meson spectrum observed
in the experiments. In particular, the mesons with $J\ge 2$
cannot be obtained from the five-dimensional gauge field.
This is, however, not a serious problem of the model.
As suggested from the original idea of string theory,
it is natural to expect that higher spin mesons 
are obtained from the massive excited states in the open
string spectrum.
The purpose of this paper is to explore this direction.
{}For related works, see
Refs.~\citen{Kruczenski:2003be}-\citen{Domokos:2009hm}. 

In this paper, we study the meson spectrum obtained from the
massive open string states.
Although it is difficult to quantize strings in the curved background
with Ramond-Ramond (RR) flux, we are able to estimate the masses of
these mesons
assuming that the 't Hooft coupling $\lambda$ is large.
The leading term in the mass formula is just
that obtained in the flat space-time, which gives the
linear Regge behavior (\ref{Regge}). Taking into account the
effect of the curved background perturbatively, we obtain
the next-to-leading terms that give corrections to
the formula (\ref{Regge}).
We also determine the quantum numbers $J^{PC}$ for these mesons
and try to identify the meson spectrum obtained from string theory
with that observed in the experiment.
We argue that it is plausible to identify
$a_2(1320)$, $b_1(1235)$, $\pi(1300)$, $a_0(1450)$, etc.,
as those obtained from the first excited massive open string states,
and $\rho_3(1690)$, $\pi_2(1670)$, etc., as the second excited states.

The organization of this paper is as follows.
First, we briefly review the model in \S \ref{review}.
In \S \ref{massive}, we analyze the meson spectrum that is obtained
from the massive open string modes.
The results are compared with the experimental data in \S
\ref{data}.
In \S \ref{summary}, we summarize our results and
discuss possible future directions. 
In Appendix \ref{higher}, we analyze the spectrum of the second
excited open string states that could be identified with excited mesons
including those with $J=3$.
In Appendix \ref{Z2}, we classify $\bZ_2$ symmetries of the system
we work in. This plays an important role in identifying
the open string states with the mesons found in the experiments.
Appendix \ref{App:RR} is devoted to a study of
the effect of RR flux into the meson mass formula.

\section{Brief review of the model}
\label{review}

In this section, we provide a brief review of the holographic QCD
proposed in Ref.~\citen{SS1} with the aim of fixing our notation and 
convention.
Here, we only describe the necessary ingredients
of the model for this paper. See Refs. \citen{SS1} and \citen{SS2} 
for more details.

The model is constructed using a system with D4/D8/\AD8-branes
in type IIA string theory compactified on an $S^1$.
To break the supersymmetry completely,
we impose the anti-periodic boundary condition on all the fermions of
the system along the $S^1$.
The radius of the $S^1$ is denoted as $\mkk^{-1}$,
although we mainly employ the unit with $\mkk=1$ in the following.
The $\mkk$ dependence can easily be recovered from the dimensional analysis.
To realize four-dimensional $SU(N_c)$ Yang-Mills theory,
we consider $N_c$ D4-branes wrapped on the $S^1$, and
$N_f$ D8-\AD8 pairs are added to obtain $N_f$ flavors of massless quarks.
The D8-branes and \AD8-branes are placed at the antipodal points
on the $S^1$ and extended along the other nine directions.

In the holographic description, the D4-branes are replaced with
the corresponding curved background considered by Witten
in Ref.~\citen{Witten_D4}.
The D8-branes are treated as probes, assuming $N_f\ll N_c$.
Then, the D8-branes and \AD8-branes are smoothly connected,
and the system becomes $N_f$ D8-branes embedded in the
Witten's D4-brane background.
The topology of the background geometry is
 $\bR^{1,3} \times \bR^{2}\times S^{4}$ and we parametrize the
 $\bR^{1,3}$ and $\bR^2$ by $x^\mu$ ($\mu=0,1,2,3$) and $(z,y)$,
respectively. The metric and dilaton configurations
can be written as
\begin{align}
&ds^2=\frac{4}{27}\lambda l_s^2 \,d\wt{s}^2\ ,
\nn\\
&d\wt{s}^2=
K(r)^{1/2} \eta_{\mu\nu}dx^{\mu}
dx^{\nu}+ K(r)^{-5/6}dr^2 + K(r)^{-1/2}r^2d\theta^2
+\frac{9}{4}K(r)^{1/6}d\Omega_4^2\ ,
\label{metric}
\end{align}
\begin{eqnarray}
 e^\phi= \frac{\lambda^{3/2}}{3\sqrt{3}\pi N_c}K(r)^{1/4}\ ,
\label{dilaton}
\end{eqnarray}
where $(r,\theta)$ is the polar coordinate
of the $\bR^2$ related to $(z,y)$ by
\begin{eqnarray}
z=r\sin \theta\ ,~~~ y=r\cos \theta\ ,
\label{yz}
\end{eqnarray}
$d\Omega_4^2$ is the line element on the unit $S^4$ and
$K(r)=1+r^2$.\footnote{
$r$ is related to $U/U_{\rm KK}$ used in Ref.~\citen{SS1} by
$(U/U_{\rm KK})^3=1+r^2$.
} The parameters
$l_s$ and $\lambda$ correspond to
the string length and 't Hooft coupling in QCD, respectively.
In addition, the RR 4-form field strength $F_4$ is proportional
to the volume form of the $S^4$ and satisfies
\begin{eqnarray}
\frac{1}{2\pi}\int_{S^4}F_4=N_c\ .
\label{RR}
\end{eqnarray}

It is important to note that the string length $l_s$ only appears in the
overall factor in the metric. When we consider
a string with tension $T=1/(2\pi l_s^2)$ embedded in this background,
the $l_s$ dependence in the string world-sheet action cancels out and
the system is equivalent to a string with tension
\begin{eqnarray}
\wt{T}\equiv\frac{2\lambda}{27\pi}
\label{tension}
\end{eqnarray}
embedded in the background,
whose metric is given by $d\wt{s}^2$ in
(\ref{metric}).
Therefore, the masses of the massive string modes
remain finite in the decoupling limit $l_s\ra 0$, implying that the
holographic dual description of QCD involves the whole massive states
of a string theory rather
than just its massless sector.
In the following, we consider strings with tension $\wt T$ embedded in
the background with metric given by $d\wt{s}^2$.
This is equivalent to setting
\begin{eqnarray}
\alpha'\equiv l_s^2 =\frac{27}{4}\lambda^{-1}\ ,
\label{alpha}
\end{eqnarray}
which is allowed because the $l_s$ dependence cancels out
anyway.\footnote{
This convention differs from that in Ref.~\citen{SS2} by a factor of 2/3 on
the right-hand side.}
In this convention, we can omit the tilde in $\wt T$ and $d\wt{s}^2$.
{}From (\ref{dilaton}) and (\ref{alpha}), we see that the loop
correction and $\alpha'$ correction in string theory correspond to
$\lambda^{3/2}/N_c$ and $1/\lambda$ corrections, respectively.
We assume that these corrections are small, which is justified when
$1\ll \lambda^{3/2}\ll N_c$.

In the D4/D8/\AD8-brane configuration (before replacing the $N_c$
D4-branes with the corresponding supergravity background),
gluons are obtained as the massless modes in the Kaluza-Klein (KK)
decomposition of the gauge field on the
D4-brane wrapped on $S^1$. The massive KK modes in the $SU(N_c)$
gauge field as well as fermions and scalar fields on the D4-brane
world-volume are artifacts of the model
and do not have any counterparts in QCD. They all have masses of order
$\mkk$, and hence, our model deviates from QCD when the energy is higher
than this scale. In principle, we should take a limit $\mkk\ra\infty$
and $\lambda\ra 0$ with the QCD scale (e.g., the rho meson mass)
kept fixed at the experimental value
to send the cutoff to infinity.
This is analogous to taking the continuum limit in lattice QCD.
The QCD scale $\Lambda_{\rm QCD}$ is related to $M_{\rm KK}$ by
$\Lambda_{\rm QCD}=f(\lambda)M_{\rm KK}$ with an unknown function
$f(\lambda)$. To take the $M_{\rm KK}\rightarrow\infty$ limit,
we have to know how the function $f(\lambda)$ behaves
in the small $\lambda$ region.
However, since our calculation can only be trusted
when $\lambda$ is large, this procedure is beyond the scope of this
paper.
Instead, we use the experimental values of two quantities
as inputs to fix $\mkk$ and $\lambda$, 
and expect that the predictions will become more accurate 
by taking into account
the $1/\lambda$ and $1/N_c$ corrections.
In Refs.~\citen{SS1} and \citen{SS2}, 
the experimental values of $\rho$ meson mass
$m_\rho|_{\rm exp}\simeq 776~{\rm MeV}$ and pion decay constant
$f_\pi|_{\rm exp}\simeq 92.4~{\rm MeV}$
are used to fix the values of the parameters $\mkk$ and $\lambda$ as
\begin{eqnarray}
\mkk\simeq 949~{\rm MeV}\ ,~~\lambda\simeq 16.6 \ .
\label{mkk}
\end{eqnarray}
With this parameter choice, various properties of hadrons predicted
in the model turned out to be in reasonably good agreement with the
experimental results. Although the cutoff scale $\mkk$ is rather low,
we can expect that the effect of the cutoff is much milder than that of
lattice QCD. In the energy scale of a few GeV,
this system can be regarded as a holographic dual of massless QCD with
extra massive fields whose masses are of order $\mkk$.
For many of the low energy quantities of hadrons we are interested in,
the main contributions are from the massless gluons and quarks, and the
effects of the extra degrees of freedom should not be large enough to
alter the order of magnitude.
The fact that our formulation does not suffer from the finite volume
effects and keeps exact Poincar\'{e} symmetry as well as chiral symmetry
also helps to reduce the error.

In this paper, we analyze the open strings attached on the D8-branes
that are interpreted as mesons in QCD.
The D8-branes are placed at $y=0$, and extended along $x^\mu$, $z$
and $S^4$ directions. The system is invariant under the $SO(5)$ isometry
that rotates the $S^4$. Since the quarks and gluons are invariant
under this $SO(5)$, only the $SO(5)$ invariant states can be identified
with the mesons in QCD. For this reason, we will focus on
the $SO(5)$ invariant field configurations.
The $SO(5)$ non-invariant states are the artifacts of the model.
Such states cannot be the bound states made only of the quarks and gluons,
but necessarily involve massive $SO(5)$ non-invariant constituents,
and hence, they are expected to be decoupled if we were able to
send the cutoff to infinity, as explained above.
In \S \ref{SO5}, we will show that there is a $\bZ_2$ symmetry
that can also be used to clean up the spectrum.

In Refs.~\citen{SS1} and \citen{SS2}, 
the massless sector of the open string is analyzed.
It was shown there that the effective theory on the D8-brane world-volume
is given by a five-dimensional YM-CS theory after
dimensional reduction on the $S^4$.
Expanding the five-dimensional gauge theory with respect to a complete
set of normalizable functions of $z$, we obtain an infinite tower of
four-dimensional meson fields, including those interpreted as pion,
$\rho$-meson, $a_1$-meson, etc.
The main goal of this paper is to extend this analysis to the massive
open string states.

One of the important quantities used in this paper
is the string tension. If we use the values (\ref{mkk}) in (\ref{alpha})
and (\ref{tension}), we obtain
$\alpha'\simeq 0.45~{\rm GeV}^{-2}$ and $T\simeq (0.59~{\rm GeV})^2$.
(See also Refs.~\citen{BrItSoYa}, \citen{HaOz}, and \citen{adscft}.)
Unfortunately, this value of string tension $T$ is considerably larger
than the value estimated in lattice gauge theory
$T|_{\rm lattice}\simeq (0.44~{\rm GeV})^2$.
(For a review, see, for example, Ref.~\citen{Teper:1997am}.)
We should not be too serious about this discrepancy, because
our analysis is not accurate enough.
It may be cured by taking into account the
$1/\lambda$ and $1/N_c$ corrections.\footnote{
See Ref.~\citen{Bigazzi:2004ze} for the calculation of the $1/\lambda$ correction
to the string tension.}
Another possible correction
may be due to the effect of quark masses. Note that the pion decay
constant is sensitive to the quark mass, and its value
in the chiral limit is about $5\sim 15\%$ smaller than the experimental value.
If we use this value to fix $\lambda$, the value of $\lambda$ will
become smaller and the predicted values of $\alpha'$ and $T$ will become
closer to the expected ones.
We will revisit this issue in \S \ref{data}.

\section{Massive string modes in the model}
\label{massive}

In this section, we consider massive open string states
in our model. Since it is not easy to quantize strings in the curved
space-time described in the previous section, we restrict our analysis
to the cases with $\lambda\gg 1$, in which the effect of the non-trivial
background can be considered perturbatively.

As explained in \S \ref{review}, we have to pick up states
that are invariant under $SO(5)$ isometry.
Then, the open string states can be regarded as particles in the
five-dimensional space-time parametrized by $x^\mu$ ($\mu=0,1,2,3$)
and $z$. We will soon show that the states that can be interpreted as
mesons in QCD should also be invariant under a $\bZ_2$ symmetry.
First, we classify the states that are invariant
under $SO(5)$ and $\bZ_2$ in \S \ref{SO5},
and then identify the parity and charge conjugation quantum numbers
in \S \ref{CP}.
We demonstrate these procedures by considering
the first excited massive string states, which include a spin-2 meson.
The results for the second excited states are given
in Appendix \ref{higher}.

The non-trivial $r$ dependence of the metric 
(\ref{metric}) is analyzed perturbatively in \S \ref{mass},
where we show that it makes the five-dimensional particles stay around
$z=0$ and they behave as four-dimensional mesons.

\subsection{Classification of open string states}
\label{SO5}

When $\lambda$ is very large, the string length is much shorter
than the scale of the curvature radii of the background,
and it is possible to approximate the space-time around a string by the
flat space-time $\bR^{1,3}\times \bR^2\times \bR^4$, where the $\bR^4$
factor denotes a local patch around a point in $S^4$.
We parametrize the flat ten-dimensional space-time
by $x^M$ ($M=0,1,\dots,9$), and also use the notation
$(z,y)=(x^4,x^5)$ and $\theta^a=x^a$ ($a=6,7,8,9$).

We are interested in the $SO(5)$ invariant states created by open
strings attached on the D8-branes placed at $y=0$.
When we approximate the background by the flat space-time,
the $SO(5)$ isometry of $S^4$ becomes $SO(4)_{6\sim 9}$ rotational and
translational symmetry acting on $\bR^4$ parametrized by $\theta^a$.
Here, $SO(n)_{A_1 \sim A_n}$ denotes the orthogonal group acting on the
$n$-dimensional space parametrized by $(x^{A_1},\dots,x^{A_n})$.
Therefore, we ought to find $SO(4)_{6\sim 9}$ invariant states with the wave
function constant along $\theta^a$ directions.
As a result, the system
is reduced to a five-dimensional space-time parametrized by $x^\mu$
($\mu=0,1,2,3$) and $z$.

To obtain the open string spectrum, it is convenient to
consider D9-branes that are related to the D8-branes by T-duality along
the $y$-direction.\footnote{
Here, we use the D9-brane picture just to learn
the open string spectrum on the D8-branes,
and do not care about the RR tadpole cancellation.
} We first compactify the $y$-direction to $S^1$,
but restrict our attention to the sector with zero winding number. 
This system is T-dual to $N_f$ D9-branes with zero momentum along
the $S^1$ direction. Here, we use the notation $\wt y=\wt x^5$ to
parametrize the T-dualized $S^1$-direction.
The background RR 4-form field strength is mapped to
an RR 5-form field strength proportional to $d\wt y\wedge d\theta^6
\wedge d\theta^7\wedge d\theta^8\wedge d\theta^9$.
Then, the $SO(5)$ invariant states
on the D8-branes are in one-to-one correspondence
with the open string states on the D9-branes that are
invariant under the $SO(4)_{6\sim 9}$ and the translational symmetry
along the five-dimensional space parametrized by $(\wt y,\theta^a)$.
There is an additional constraint we have to impose to obtain the states
corresponding to the mesons in QCD.
In fact, as we will show in Appendix \ref{Z2},
a $\bZ_2$ action generated by
\begin{eqnarray}
 (\wt y, x^9)\ra (-\wt y,-x^9)
\label{y9}
\end{eqnarray}
is a symmetry of the whole system,
under which quarks and gluons are invariant.
Therefore, the mesons in QCD should correspond to the open string
states that are also invariant under this $\bZ_2$.\footnote{
This $\bZ_2$ is T-dual of the ``$\tau$-parity'' used
in Ref.~\citen{BrMaTa} to classify the closed string states
corresponding to glueballs in QCD.
The invariance under this $\bZ_2$ implies that the massless
scalar field on the D8-brane world-volume
considered in \S 4.2 of Ref.~\citen{SS1}
cannot be interpreted as scalar mesons in QCD.
See Appendix \ref{Z2} for further comments.
}
We call this $\bZ_2$ symmetry ``$\tau$-parity'', following 
Ref.~\citen{BrMaTa}.

Massive particles created by the strings attached
on the D9-brane can be classified by representations of
$SO(9)_{1\sim 9}$ that is the little group for the ten-dimensional
massive particle. We first classify the massive string
modes by the representations of $SO(9)_{1\sim 9}$, decompose them by the
representations of $SO(5)_{1\sim \wt 5}\times SO(4)_{6\sim 9}$
subgroup, and extract the $SO(4)_{6\sim 9}$ invariant components.
Then, we pick up the components that are invariant under the
$\tau$-parity (\ref{y9}).

Let us examine the first excited states explicitly following the procedure
explained above. We use the light-cone gauge quantization in the NSR
formalism to quantize the open strings on the D9-brane in the T-dualized
picture. (See, for example, Ref. \citen{GSW}.)
We take $x^0\pm x^1$ as the light-cone directions.
Since the states in R-sector cannot be invariant under $SO(4)_{6\sim 9}$,
we consider only the NS-sector.
The physical states are constructed by acting
with an arbitrary number
of the bosonic creation operators $\alpha^I_{-n}$
 ($I=2\sim 9$\,; $n=1,2,3,\dots$)
and an odd number of fermionic creation operators $\psi^J_{-r}$
 ($J=2\sim 9$\,; $r=1/2,3/2,5/2,\dots$)
on the NS-vacuum $\ket$ as
\begin{eqnarray}
\alpha^{I_{1}}_{-n_1}\cdots \alpha^{I_k}_{-n_k}
\psi^{J_{1}}_{-r_1}\cdots \psi^{J_{2l+1}}_{-r_{2l+1}}
\ket
\label{states}
\end{eqnarray}
with non-negative integers $k$ and $l$.
Here, the NS-vacuum $\ket$ is an eigenstate of the momentum operator
with the momentum $p$ satisfying the mass
shell condition
\begin{eqnarray}
-p^2 = \frac{N}{\alpha'}
=\frac{4}{27}\lambda N
\equiv m_0^2\ ,
\label{massshell}
\end{eqnarray}
where we have used (\ref{alpha}) and
\begin{eqnarray}
N\equiv \sum_{i=1}^{k} n_i+\sum_{i=1}^{2l+1} r_i-\frac{1}{2}\ .
\label{N}
\end{eqnarray}
Here,
the momentum $p$ can take non-zero values only in the five-dimensional
components $p=(p^\mu,p^z)$ ($\mu=0,\cdots,3$),
because we impose the translational invariance along $(\wt y,\theta^a)$.

The first excited state with $N=1$ is
\begin{eqnarray}
 \psi_{-3/2}^J\ket\ ,~~
 \alpha_{-1}^I \psi_{-1/2}^J\ket\ ,~~
 \psi_{-1/2}^{J_1}\psi_{-1/2}^{J_2}\psi_{-1/2}^{J_3}\ket\ ,
\label{N=1}
\end{eqnarray}
which gives vector representation ($\fnd_{\,8}$), rank-2 tensor
representation ($\fnd_{\,8}\otimes\fnd_{\,8}$), and rank-3 anti-symmetric
tensor representation ($\,\Tasym_{\,56}$) of $SO(8)_{2\sim 9}$,
respectively. Here, the subscript of each Young tableau represents
the dimension of the representation.
The rank-2 tensor representation is reducible and it can be decomposed
to singlet ($\ul 1$), symmetric traceless representation ($\sym_{\,35}$), and
anti-symmetric representation ($\,\asym_{\,28}$).
Although the $SO(9)_{1\sim 9}$ symmetry is not manifest in the 
light-cone gauge,
it is not difficult to see that these multiplets are obtained from
$\,\Tasym_{\,84}\oplus\,\sym_{\,44}$ of $SO(9)_{1\sim 9}$.
The decomposition of these representations with respect to
the representation of
$SO(5)_{1\sim \wt 5}\times SO(4)_{6\sim 9}\subset SO(9)_{1\sim 9}$
is given by
\begin{align}
\sym_{\,44}
& =(\sym_{\,14},\ul 1)\oplus(\ul 1,\sym_{\,9})
   \oplus(\fnd_{\,5},\fnd_{\,4})\oplus(\ul 1,\ul 1)\ ,
\nn\\
\Tasym_{\,84}
& =(\,\Tasym_{\,10},\ul 1)\oplus(\,\asym_{\,10},\fnd_{\,4})
   \oplus(\fnd_{\,5},\asym_{\,6})\oplus(\ul 1,\Tasym_{\,4})\ .
\end{align}
Therefore, the $SO(4)_{6\sim 9}$ invariant states are
\begin{eqnarray}
\ul1\,\oplus\,\Tasym_{\,10}\oplus\,\sym_{\,14} 
\label{6dimN=1}
\end{eqnarray}
of $SO(5)_{1\sim \wt 5}$.

Thus far, we have considered the first excited massive states for the open
strings attached on the D9-brane, which is T-dual of the D8-brane.
The field content for the D8-brane is obtained from the
dimensional reduction of that for the D9-brane.
Then, the five-dimensional fields corresponding to the first excited states
on the D8-brane can be obtained by the dimensional reduction of 
the six-dimensional fields listed in (\ref{6dimN=1}), that is,
the scalar field $\varphi$, rank-3 anti-symmetric tensor
field $A_{\alpha\beta\gamma}$,
and traceless symmetric tensor field $h_{\alpha\beta}$
($\alpha,\beta,\gamma=1,\dots,5$).
The components that are invariant
under the $\tau$-parity (\ref{y9}) are
\begin{eqnarray}
\varphi\ ,~~A_{MNP}\ ,~~h_{MN}\ ,~~h_{yy}
\label{5dim}
\end{eqnarray}
with $M,N,P=1,2,3,z$.

Higher excited states can be constructed in a similar manner.
We present the results for the second excited states
in Appendix \ref{higher}.
One of the important properties worth mentioning here is that,
as it has been well-known from the early days of string theory,
the spectrum exhibits linear Regge behavior. 
Namely, the highest spin $J$ for each
mass level is $J=N+1=1+\alpha' m_0^2$.
In \S \ref{mass}, we show that
the masses of four-dimensional mesons are
modified from this linear behavior.

\subsection{Mass spectrum}
\label{mass}

To obtain four-dimensional meson fields from the
five-dimensional fields obtained in \S \ref{SO5},
we expand the five-dimensional fields using
a complete set of normalizable functions of $z$ that is
chosen to extract mass eigenstates.
The idea is the same as the usual Kaluza-Klein (KK) decomposition.
Although the space in the $z$ direction is non-compact,
the non-trivial $z$ dependence of the metric induces
a potential that prevents particles from moving away to infinity
and the system  effectively becomes four-dimensional.

For example, the five-dimensional
scalar field $\varphi$ is expanded as
\begin{eqnarray}
\varphi(x^\mu,z)=\sum_{n=0}^\infty\varphi^{(n)}(x^\mu)\phi_n(z)\ ,
\label{expand}
\end{eqnarray}
where $\{\phi_n(z)\}_{n\ge 0}$ is a complete set of
normalizable functions of $z$ to be determined
and $\varphi^{(n)}(x^\mu)$ denote the four-dimensional meson fields.
The field equation corresponding to the mass shell condition
(\ref{massshell}) in the flat space-time limit is
\begin{eqnarray}
\left(
\eta^{\mu\nu}\del_\mu\del_\nu+\del_z^2-m_0^2
\right)\varphi(x^\mu,z)=0\ ,
\label{flateom}
\end{eqnarray}
where $\mu,\nu=0,\dots,3$ and $m_0^2=N/\alpha'$ as defined in
(\ref{massshell}).
Here, we ignore the interaction terms assuming that the string coupling
$g_s\simeq \lambda^{3/2}/N_c$ (see (\ref{dilaton})) is small.
In the following,
we claim that the leading correction to this equation
in the curved space-time can be included by just replacing
$\eta^{\mu\nu}$ in (\ref{flateom}) by the curved metric
$g^{\mu\nu}(z)=K(z)^{-1/2}\eta^{\mu\nu}$ in (\ref{metric}) as
\begin{eqnarray}
\left(
g^{\mu\nu}(z)\del_\mu\del_\nu+\del_z^2-m_0^2
\right)\varphi(x^\mu,z)=0
\label{curved1}
\end{eqnarray}
for large $\lambda$. More precisely, for the warped geometry with
$g_{\mu\nu}(z)=f(z)\,\eta_{\mu\nu}$ and $f(z)=1+c\, z^2+ O(z^4)$
(in our case, $f(z)=K(z)^{1/2}$ and $c=1/2$),
the terms we should keep in (\ref{curved1}) $\times f(z)$ are
\begin{eqnarray}
\left(
\eta^{\mu\nu}\del_\mu\del_\nu+\del_z^2-m_0^2(1+c\,z^2)
\right)\varphi(x^\mu,z)=0\ ,
\label{eom}
\end{eqnarray}
and all the other terms are sub-leading for large $\lambda$.
This statement holds not only for the scalar fields but also for any
tensor fields.

To see that the other possible terms in the equations
of motion (\ref{eom}) can be neglected,
it is convenient to rescale $z$ as
\begin{eqnarray}
w=\lambda^{1/4}z
\label{rescale}
\end{eqnarray}
and treat $w$ as an $\cO(1)$ variable.
The motivation for this rescaling
is that, as we will soon show, the typical width of the wave functions
in the $z$-direction turns out to be of $\cO(1)$
in terms of the rescaled coordinate $w$. Recall that $m_0^2$ defined in
(\ref{massshell}) is proportional to $\lambda$, and hence,
$\eta^{\mu\nu}\del_\mu\del_\nu$ in (\ref{eom}) should be considered to
be of order $\lambda$. Because of the rescaling (\ref{rescale}),
 $\del_z^2=\lambda^{1/2}\del_w^2$ and $m_0^2 z^2
 =(m_0^2\lambda^{-1/2})\,w^2$ in
(\ref{eom}) are of order $\lambda^{1/2}$. These terms give the
leading corrections to the flat space-time limit.
We keep up to $\cO(\lambda^{1/2})$ terms and neglect the
smaller contributions in the field equation assuming $\lambda$ is large.
It is easy to see that the contributions from
possible $\cO(z^2)$ terms in front of $\del_z^2$ and $\cO(z^4)$ terms in
$m_0^2f(z)$ in (\ref{eom}) are at most of $\cO(1)$ and can be
neglected. Since (\ref{flateom}) is exact in the flat space-time limit,
all the other corrections should involve the derivatives of the background
metric, dilaton, and RR fields. The metric and dilaton
configurations in (\ref{metric}) and (\ref{dilaton}) are independent of
$x^\mu$, and the $z$-derivatives of metric and dilaton ($\del_z g_{MN}$
and $\del_z \phi$) are of $\cO(\lambda^{-1/4})$. Therefore, the terms with
derivatives of metric and dilaton field cannot contribute to the
$\cO(\lambda^{1/2})$ terms.
This implies that, for example, we do not have to replace the derivatives 
in (\ref{curved1}) with the covariant derivatives, since the terms with 
Christoffel symbols
are sub-leading. The corrections involving the curvature of the
space-time are also negligible.
One might think that the higher derivative terms
may have larger contributions, since $\del_z\sim\cO(\lambda^{1/4})$.
However, since each derivative is accompanied by
$l_s\sim\lambda^{-1/2}$, the higher derivative terms can be
neglected as well.
The effect of the RR field is trickier. We show in
Appendix \ref{App:RR} that the corrections to the equations of motion
(\ref{eom}) due to the background RR field (\ref{RR})
can also be neglected within our approximation.

{}From (\ref{eom}), the masses of the four-dimensional meson fields
are obtained as the eigenvalues of the eigenequation
\begin{eqnarray}
\left(-\del_z^2+m_0^2(1+c\,z^2)\right)\phi_n(z)
=M_n^2\,\phi_n(z)\ .
\label{eigen}
\end{eqnarray}
The eigenfunctions $\{\phi_n(z)\}_{n\ge0}$ form
a complete set, and are used to expand the five-dimensional field
as in (\ref{expand}), with the eigenvalue $M_n^2$
interpreted as the mass squared
of the $n$th meson field $\varphi^{(n)}(x^\mu)$.
The eigenequation (\ref{eigen}) is the same as the Schr\"odinger
equation of the harmonic oscillator 
so that
\begin{align}
M_n^2&=m_0^2+2m_0\sqrt{c}\left(n+\frac{1}{2}\right)\nn\\
&=\frac{N}{\alpha'}+\sqrt{\frac{2N}{\alpha'}}
\left(n+\frac{1}{2}\right) \mkk
\label{Mn1}
\end{align}
with $n=0,1,\dots$, where we have used $c=1/2$ and
(\ref{massshell}), and recovered $\mkk$ by the dimensional analysis. 
The eigenfunctions are
$\phi_n(z)\propto H_n(\xi)\exp(-\xi^2/2)$, where
$\xi=(m_0^2c)^{1/4}z$ and  $H_n(\xi)$ is the $n$th Hermite polynomial,
and they satisfy $\phi_n(-z)=(-1)^n\phi_n(z)$.

It would be instructive to derive the formula (\ref{Mn1})
in a more direct way. The action for a particle of mass $m_0$
placed on the D8-brane is
\begin{eqnarray}
S= -m_0\int dt\,\sqrt{-g_{tt}}
= -m_0 \int dt\,K(z)^{1/4} \ ,
\end{eqnarray}
which implies that this particle is trapped around $z=0$ with
a potential
\begin{eqnarray}
V(z)\equiv m_0 K(z)^{1/4}
\simeq m_0 \left(1+\frac{z^2}{4}\right)+\cO(z^4)\ .
\label{V}
\end{eqnarray}
Approximating the potential $V(z)$ by a harmonic oscillator,
the energy eigenvalues are obtained as
\begin{eqnarray}
M_n\simeq m_0+\frac{1}{\sqrt{2}}\left(n+\frac{1}{2}\right)\mkk\ .
\label{Mn2}
\end{eqnarray}
This result consistently reproduces (\ref{Mn1})
up to $\cO(\lambda^{1/2})$ terms.

\subsection{Parity and charge conjugation}
\label{CP}

In Ref.~\citen{SS1}, it was shown that the parity transformation in QCD
corresponds to flipping the sign of the spatial coordinates in
the five-dimensional space-time:
\begin{eqnarray}
(x^0,x^1,x^2,x^3,z) \ra (x^0,-x^1,-x^2,-x^3,-z)\ ,
\label{parity}
\end{eqnarray}
and the charge conjugation is a $\bZ_2$ transformation acting on the
massless five-dimensional field $A_M(x^\mu,z)$ ($M=0,1,2,3,z$ ;
$\mu=0,1,2,3$) on the D8-brane as
\begin{eqnarray}
A_\mu (x^\mu,z) \ra -A_\mu^T (x^\mu,-z) \ ,~~
A_z (x^\mu,z) \ra A_z^T (x^\mu,-z)\ .
\label{C1}
\end{eqnarray}
The parity transformation (\ref{parity}) works for the massive
sector in the same manner. 
However, the charge conjugation (\ref{C1}) is given only for the
massless gauge field and we need to know how it acts on
the massive string modes.
Because the charge conjugation interchanges the
left-handed and right-handed components of the quark field, we expect
that the coordinate $z$ should be mapped to $-z$. In addition,
it relates a field with its transpose as in (\ref{C1}), implying
that
the orientation of the open string should be flipped by the charge
conjugation. Therefore, the charge conjugation corresponds to
an orientifold action that involved the reflection of $z$. 
An orientifold
action consistent with our brane configuration was found in
Ref.~\citen{ImSaSu}, in which O6-planes are added to the system to obtain a
holographic description of $O(N_c)$ and $USp(N_c)$ QCD.
It is shown that the $\bZ_2$ symmetry corresponding to the charge 
conjugation is the orientifold action associated with an O$6^+$-plane,
which is defined by
\begin{eqnarray}
(z,x^8,x^9)\ra (-z,-x^8,-x^9)
\label{O6}
\end{eqnarray}
together with world-sheet parity transformation.\footnote{
The orientifold action associated with an O6-plane is
defined by $I\Omega (-1)^{F_L}$ for closed strings, where
$I$ is an involution that flips three spatial coordinates,
$\Omega$ is the world-sheet parity transformation, and $F_L$ is the
left-moving space-time fermion number.
}

Let $\cC$ be the generator of this $\bZ_2$ represented on
the open string states.
 The action of $\cC$ on states in the light-cone gauge
(\ref{states}) can be read from the relations:
\begin{align}
&\cC\alpha^{\parallel}_{-n}\cC^{-1}=(-1)^n \alpha^{\parallel}_{-n}\ ,~~
\cC\alpha^{\perp}_{-n}\cC^{-1}=-(-1)^n \alpha^{\perp}_{-n}\ ,
\nn\\
&\cC\psi^{\parallel}_{-r}\cC^{-1}=e^{ir\pi}\psi^{\parallel}_{-r}\ ,~~
\cC\psi^{\perp}_{-r}\cC^{-1}=-e^{ir\pi}\psi^{\perp}_{-r}\ ,
\nn\\
&\cC \left|\, 0;A,B\,\right>=i\left|\, 0;B,A\,\right>\ ,
\label{C}
\end{align}
where $\parallel\,=2,3,6,7$ ; $\perp\,=z,y,8,9$, and
 $\left|0;A,B\right>$ is the NS-vacuum with Chan-Paton indices
$A,B$. Since we are interested in the $SO(4)_{6\sim 9}$ invariant
states, we can also use $\bZ_2$ defined by (\ref{C}) with
$\parallel\,=2,3,6,7,8,9$ and $\perp\,=z,y$ to examine the charge
conjugation parity. It acts on the $SO(4)_{6\sim 9}$ invariant
states of the form (\ref{states}) as  $(-1)^{N+n_y+n_z+1}$,
where $N$ is the excitation level defined in (\ref{N}), 
$n_y$ and $n_z$ are the numbers of $y$ and $z$ indices, respectively.
{}Furthermore, it is accompanied by interchanging the Chan-Paton indices
and mapping $z\ra -z$ in the argument of the wave function. 

It is easy to check that the charge conjugation defined in (\ref{C})
yields (\ref{C1}) for the massless states $\psi_{-1/2}^M\ket$.
For the first excited states obtained in (\ref{5dim}),
the charge conjugation $\cC$ acts as
\begin{align}
&A_{ijk}(x^\mu,z)\ra A_{ijk}^T(x^\mu,-z) \ ,~~
A_{ijz}(x^\mu,z)\ra -A_{ijz}^T(x^\mu,-z) \ ,\nn\\
&h_{ij}(x^\mu,z)\ra h_{ij}^T(x^\mu,-z) \ ,~~
h_{iz}(x^\mu,z)\ra -h_{iz}^T(x^\mu,-z) \ ,\nn\\
&h_{zz}(x^\mu,z)\ra h_{zz}^T(x^\mu,-z) \ ,~~
h_{yy}(x^\mu,z)\ra h_{yy}^T(x^\mu,-z) \ ,~~
\varphi(x^\mu,z)\ra \varphi^T(x^\mu,-z) \ ,
\label{C2}
\end{align}
where $i,j,k=1,2,3$.

{}Four-dimensional meson fields
are obtained by expanding the five-dimensional fields
as in (\ref{expand}). The parity and charge conjugation properties
can be read from (\ref{parity}) and (\ref{C2}), together with the fact
that the mode functions satisfy $\phi_n(-z)=(-1)^n\phi_n(z)$.
The spin $J$, parity $P$, and charge conjugation parity $C$
for the lightest mesons with $n=0$ in the first excited
massive string states (\ref{5dim}) are as follows:
\begin{eqnarray}
\begin{array}{c|ccccccccccc}
&h_{ij}^{(0)}&h_{i z}^{(0)}&h_{zz}^{(0)}&h_{yy}^{(0)}&
A_{ijk}^{(0)}&A_{ijz}^{(0)}&\varphi^{(0)}\\
\hline
J^{PC}&2^{++}&1^{+-}&0^{++}&0^{++}&0^{-+}&
1^{--}&0^{++}
\end{array}
\ .
\label{hAphi}
\end{eqnarray}
{}For the second lightest modes with $n=1$,
$P$ and $C$ are all flipped,
compared with those with $n=0$.

\section{Comparison with the experimental data}
\label{data}

\begin{figure}[ht]
\begin{center}
\includegraphics[width=15cm]{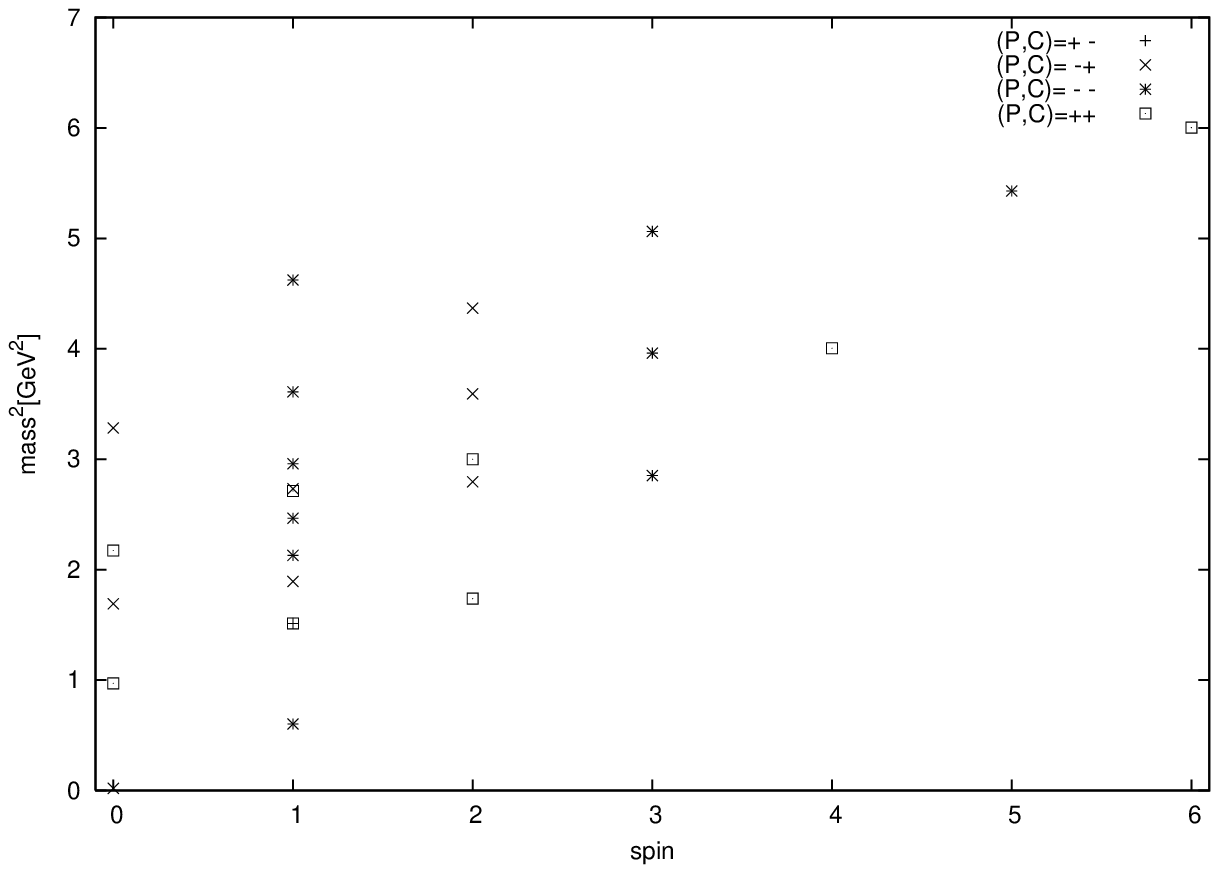}
\end{center}
\refstepcounter{figure}
\begin{center}
{\small
Fig.\thefigure~
Mass squared against spin of the isovector mesons in Ref. \citen{PDG}.}
\label{exp}
\end{center}
\end{figure}

Here, we try to compare our results with the experimental data.
Since our model only contains massless quarks, we set $N_f=2$
and focus on the light unflavored mesons.
In particular, we consider the isovector mesons,
because isoscalar mesons could be mixed with glueballs.
The isovector mesons seen in the meson summary table in Ref.~\citen{PDG} are
listed in Table \ref{table1}.
A plot of mass squared against spin for these mesons is shown
in Fig. \ref{exp}.
\begin{center}
\vspace*{.5ex}
\begin{tabular}[t]{l||l|l|l|l|l|l}
\hline
$0^{-+}(\pi)$&~\,135&1300&1800&&&\\
\hline
$0^{++}(a_{0})$&~\,980&1450&&&&\\
\hline
$1^{--}(\rho)$ &~\,770 & 1450 & $1570^{\triangle}$ & 1700 & $1900^{\triangle}$ 
&$2150^{\triangle}$ \\
\hline
$1^{++}(a_{1})$ & 1260 & $1640^{\triangle}$ &&&& \\
\hline
$1^{+-}(b_{1})$ & 1235 &
 &&&& \\
\hline
$1^{-+}(\pi_{1})$& 1400 & 1600 &&&& \\
\hline
$2^{++}(a_{2})$& 1320 & $1700^{\triangle}$ &&&& \\
\hline
$2^{-+}(\pi_{2})$& 1670 & 1880 & $2100^{\triangle}$ &&& \\
\hline
$3^{--}(\rho_{3})$& 1690 & $1990^{\triangle}$ & $2250^{\triangle}$ &&& \\
\hline
$4^{++}(a_{4})$& 2040 &&&&& \\
\hline
$5^{--}(\rho_{5})$& $2350^{\triangle}$ &&&&& \\
\hline
$6^{++}(a_{6})$& $2450^{\triangle}$ &&&&& \\
\hline
\end{tabular}
\\
\vspace*{2ex}
\refstepcounter{table}
\label{table1}
\parbox{15cm}{\small
Table \thetable~ 
Isovector mesons from meson summary table in Ref.~\citen{PDG}.
The left row denotes $J^{PC}$ and each three- or four-digit 
number on the table represents the approximate mass used as a part
of the name of the meson in units of MeV.
The superscript $\triangle$ is assigned for the mesons that are not
regarded as being established.
}
\end{center}

\begin{figure}[ht]
\begin{center}
\includegraphics[width=13cm]{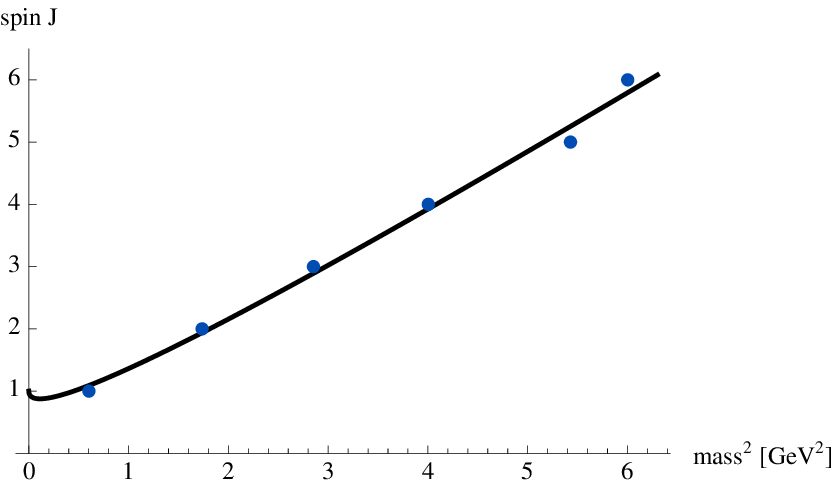}
\end{center}
\refstepcounter{figure}
\begin{center}
\parbox{15cm}{\small
Fig.\thefigure~
A plot of (\ref{J2}) with $\alpha'=1.1~{\mbox{GeV}^{-2}}$.
The dots represent mesons in the $\rho$-meson trajectory.}
\label{ReggePlot}
\end{center}
\end{figure}

{}From Table \ref{table1} and Fig.\ref{exp},
we observe that the mass squared $M^2$ against spin $J$ of
$\rho(770)$, $a_2(1320)$, $\rho_3(1690)$, $a_4(2040)$, $\rho_5(2350)$,
and $a_6(2450)$ lies on a linear trajectory satisfying
\begin{eqnarray}
J =\alpha_0+\alpha'M^2
\label{J1}
\end{eqnarray}
with $\alpha_0|_{\rm exp}\simeq 0.53$ and
 $\alpha'|_{\rm exp}\simeq 0.88~ {\rm GeV^{-2}}$.
The mesons in this $\rho$-meson trajectory 
are the lightest mesons with a given spin, and hence,
they should correspond to the states with $N=J-1$ and
$n=0$ in (\ref{Mn1}). 
The mass formula (\ref{Mn1}) implies
\begin{eqnarray}
J\simeq 1+\alpha'M_{n=0}^2-\frac{\alpha'}{\sqrt{2}}
 \mkk M_{n=0}+\cO(1/\lambda)
\label{J2}
\end{eqnarray}
for these mesons except for $\rho(770)$.
The lightest vector meson
$\rho(770)$ is obtained from the massless sector with $N=0$, for which
we cannot use the mass formula (\ref{Mn1}).
As mentioned in \S \ref{SO5}, the leading terms in (\ref{J2})
reproduce the linear trajectory (\ref{J1}) with $\alpha_0=1$.
The subleading $\cO(\lambda^{-1/2})$ term in (\ref{J2})
gives the deviation from it.
Note that a similar non-linear behavior is found in the analysis of
spinning strings in Refs.~\citen{PandoZayas:2003yb} and 
\citen{Bigazzi:2004ze}.
Unfortunately, as discussed in \S \ref{review}, the value of
$\alpha'$ evaluated from (\ref{alpha}) and (\ref{mkk}) is not good
enough to fit the experimental data.
Given the fact that the correction to $\alpha'$ in
(\ref{alpha}) and (\ref{mkk}) should be large, it may be interesting to
see if the observed meson masses can be obtained by adjusting the value
of $\alpha'$. In fact, if we set $\alpha'\simeq 1~{\rm GeV^{-2}}$,
the masses predicted by the formula (\ref{J2})
are in very good agreement with those of
$a_2(1320)$, $\rho_3(1690)$, $a_4(2040)$, $\rho_5(2350)$,
and $a_6(2450)$. (See Fig.~\ref{ReggePlot}.)
Here, the ${\cal O}(\lambda^{-1/2})$ term in (\ref{J2}) plays a crucial role,
because the linear Regge trajectory with $\alpha_0=1$ can never
give a good fit to the experimental data.

In the following, we argue the possible identification of the
open string states in (\ref{hAphi}) and (\ref{N=2}) with the mesons
listed in Table \ref{table1}.
Since there is an ambiguity in the value of $\alpha'$,
we do not use the values of meson masses predicted using (\ref{Mn1})
explicitly for our purpose.
Instead, we rely on the property of the mass formula (\ref{Mn1}) that
the mesons identified with the open string states with the
same $(N,n)$ should be nearly degenerate.
Noting the fact that the mesons
$a_2(1320)$, $\rho_3(1690)$, $a_4(2040)$, $\rho_5(2350)$,
and $a_6(2450)$ are naturally identified with
string modes with $N=1,2,3,4,5$ and $n=0$,
we expect that the mesons identified
with open string states in (\ref{hAphi}) and (\ref{N=2})
should have masses close to that of $a_2(1320)$
($m_{a_2}\simeq 1318$ MeV)
and $\rho_3(1690)$ ($m_{\rho_3}\simeq 1689$ MeV), respectively.

Let us begin by reconsidering the identification of the massless
($N=0$) modes studied in Refs.~\citen{SS1} and \citen{SS2}.
According to Refs.~\citen{SS1} and \citen{SS2}, 
the massless five-dimensional gauge field
on the D8-brane produces a pseudo-scalar meson ($J^{PC}=0^{-+}$), vector
mesons ($J^{PC}=1^{--}$), and axial-vector mesons ($J^{PC}=1^{++}$). The
predicted masses of the low-lying states are listed in
Table \ref{rhoa1}.
\begin{center}
\begin{tabular}[t]{l||l|l|l}
\hline
$0^{-+}(\pi)$&~~~~\,0&&\\
\hline
$1^{--}(\rho)$ &\,[776] & 1607 & 2435 \\
\hline
$1^{++}(a_{1})$ & 1189 & 2024 & 2849 \\
\hline
\end{tabular}
\\
\vspace*{2ex}
\refstepcounter{table}
\parbox{15cm}{\small
Table \thetable~
Mesons obtained from the massless sector in Refs.\citen{SS1} and 
\citen{SS2}. Here, we have
used the rho meson mass $m_\rho\simeq 776~{\rm MeV}$ as an input to fix
 the value of $\mkk$ as in (\ref{mkk}).
}
\label{rhoa1}
\end{center}
The massless pseudo-scalar meson in Table \ref{rhoa1} is interpreted as
pion. It appears as a massless Nambu-Goldstone particle
associated with the spontaneous chiral symmetry breaking,
as shown in Ref.~\citen{SS1}. We could not reproduce the mass of the pion,
simply because our model corresponds to QCD with massless quarks.
It would be worth emphasizing that the vector and axial-vector mesons
obtained from the massless five-dimensional gauge field are all massive,
despite the fact that the Regge intercept for the Regge trajectory
obtained in the flat space-time limit is $\alpha_0=1$.
In Refs.~\citen{SS1} and \citen{SS2},
the lightest vector and axial-vector mesons in Table \ref{rhoa1}
are interpreted as $\rho(770)$ and $a_1(1260)$, respectively.
Since we do not see any axial-vector mesons with $J^{PC}=1^{++}$
in (\ref{hAphi}), it is indeed plausible to identify $a_1(1260)$ as a
meson obtained from the massless string modes. The predicted mass 
(1189 MeV) is
very close to the observed value ($1230\pm 40$ MeV) for $a_1(1260)$.
The interpretation of the heavier modes in Table \ref{rhoa1} is less clear.
Among the vector mesons in Table \ref{table1}, $\rho(1570)$ has the
closest mass to the second lightest vector meson in Table \ref{rhoa1}.
However, since our approximation is not good enough to make a definite
identification, $\rho(1450)$ and $\rho(1700)$ are also good candidates.

Our result for the first excited massive string modes
in (\ref{hAphi}) predicts that there should be
mesons with $J^{PC}=2^{++}$, $1^{--}$, $1^{+-}$, $0^{++}$, and
$0^{-+}$ with approximately the same masses.
The candidates for the mesons with $J^{PC}=2^{++}$,
 $1^{+-}$, $0^{++}$, and $0^{-+}$ are $a_2(1320)$,
 $b_1(1235)$, $a_0(1450)$, and $\pi(1300)$, respectively.
These mesons cannot be interpreted as those from the massless open
string modes in Table \ref{rhoa1} and it is nice to have candidates
of these mesons in the first excited massive string states. The masses
of these mesons are reasonably close to each other.
Note that $a_0(980)$, which is the lightest meson with $J^{PC}=0^{++}$
in Table \ref{table1}, is considered to be a four-quark state or
two-meson resonance.
 (See the section of ``Non-$q\ol q$ Candidates'' in Ref.~\citen{PDG}.)
Since the interaction among $q\ol q$ mesons vanishes in the
large $N_c$ limit, the four-quark states cannot appear as stable
bound states in large $N_c$ QCD.
Therefore, we do not interpret $a_0(980)$ as one of the $J^{PC}=0^{++}$
states in (\ref{hAphi}).
According to (\ref{hAphi}), two more $0^{++}$ states are predicted.
We are not sure how to interpret these modes.
The most plausible candidate for the $J^{PC}=1^{--}$ state in
(\ref{hAphi}) is $\rho(1450)$. However, since we also have
 $J^{PC}=1^{--}$ states in the $N=0$ sector as discussed above,
this identification is rather ambiguous.

The spectrum of second excited massive open string modes
with $N=2$ is analyzed in Appendix \ref{higher}. Among those summarized
in (\ref{N=2}), the states with $J^{PC}=3^{--}$ and $2^{-+}$
do not appear in the spectrum of the massless $(N=0)$ and
the first excited $(N=1)$ states. 
Therefore,
they should be interpreted as the lightest mesons with $J^{PC}=3^{--}$
and $2^{-+}$  in Table \ref{table1}, that is,
$\rho_3(1690)$ and $\pi_2(1670)$, respectively.
It is encouraging to note that
the masses of these two mesons are very close to each other.
$\pi_2(1880)$ could also be a candidate for the $2^{-+}$ meson
in (\ref{N=2}). 
Although the experimental evidence for the existence is 
insufficient, 
the second lightest mesons with $J^{PC}=1^{++}$ and $2^{++}$
in Table \ref{table1} ($a_1(1640)$ and $a_2(1700)$) have
approximately the same mass as $\rho_3(1690)$, indicating
that these mesons could be interpreted as those in (\ref{N=2}).
In (\ref{N=2}), we find six $J^{PC}=0^{-+}$ states. Although the
interpretation of the degeneracy is unclear, a natural candidate
for one of these states is $\pi(1800)$,
since we have interpreted $\pi(1300)$ as one of the $N=1$ states above.
There are also a lot of $J^{PC}=1^{--}$ states in (\ref{N=2}),
which could be interpreted as the $1^{--}$ mesons in Table \ref{table1}.
For example, $\rho(1700)$ is nearly degenerate with $\rho_3(1690)$
and it is tempting to interpret it as one of the $N=2$ states,
although we cannot exclude the possibility that $\rho(1700)$ is
one of the $1^{--}$ states found in the spectrum of $N=0$ and
$N=1$ states as discussed above.
The mesons with $J^{PC}=1^{-+}$ can be obtained from the states with
$(N,n)=(1,1)$ and $(N,n)=(2,0)$. $\pi_1(1400)$ and $\pi_1(1600)$ are the
candidates in Table \ref{table1}.
However, the mass of $\pi_1(1400)$ ($1351\pm 30$ MeV) is rather close
to those of the $(N,n)=(1,0)$ states considered above,
and it seems unclear if $\pi_1(1400)$ should be interpreted as one of the
$(N,n)=(1,1)$ states or $(N,n)=(2,0)$ states.
In fact, it is known that mesons with $J^{PC}=1^{-+}$ are exotic states
that cannot be obtained as $q\ol q$ bound states.
 $\pi_1(1400)$ and $\pi_1(1600)$ are thus regarded as four-quark states
or hybrid states ($q\ol q$ pairs bound by excited gluons). 
There are some works suggesting that $\pi_1(1400)$ is a four-quark state
and $\pi_1(1600)$ is a candidate
of a hybrid meson \cite{Iddir:2000yb,Chung:2002fz}. If this is the case,
$\pi_1(1400)$ is not interpreted as the open string states considered in
this paper. On the other hand, the hybrid mesons exist as narrow
resonances in large $N_c$ QCD \cite{Cohen} and should appear
as open string states in our analysis.
The mass of $\pi_1(1600)$ ($1662^{+15}_{-11}$ MeV) is also
close to that of $\rho_3(1690)$ and
it is plausible to interpret it as the $J^{PC}=1^{-+}$ state
with $(N,n)=(2,0)$ in (\ref{N=2}).

It might look strange that there are
no clear candidates in Table \ref{table1}
to be identified with the open string massive states with $n\ge 1$.
{}For example, the states with $N=n=1$ include
those with $J^{PC}=0^{--}$, $0^{+-}$, and $2^{--}$, which are not found
in the experiments. This fact may suggest that the states with $n\ge 1$
are relatively heavy or correspond to wide resonances.
This may again indicate that $\pi_1(1600)$ corresponds to a state with
$(N,n)=(2,0)$ rather than that with $(N,n)=(1,1)$.
If we naively use our mass formula (\ref{Mn2}) with the value
of $\mkk$ in (\ref{mkk}), we obtain $M_{n+1}-M_{n}\simeq 671~{\rm MeV}$,
which predicts that the masses of the states with $(N,n)=(1,1)$ are
approximately $2000~{\rm MeV}$.
It is, however, unclear to what extent we can trust
this value, since the contribution from the second term in (\ref{mkk})
is comparable to or larger than that from the first term for $n\ge 1$.

There are some other states in (\ref{N=2}) that do not
have plausible candidates in Table \ref{table1}.
{}For example, our result (\ref{N=2}) predicts that there are mesons with
$J^{PC}=0^{++}, 1^{+-}$, and $2^{--}$
that have masses close to $\rho_3(1690)$. It
would be interesting if such mesons are found in future experiments.

\section{Summary and discussion}
\label{summary}

In this paper, we have studied the higher excited meson spectrum
using a holographic description of QCD proposed in Ref.~\citen{SS1}.
One of the motivations of this investigation is to discuss
whether or not the gauge/string duality holds beyond the massless sector
in the holographic description.
In fact, the higher excited mesons originate from 
the massive modes of open strings that end on probe D8-branes.
We derived a mass formula for these mesons by quantizing the open string
perturbatively.
It was found that the mass formula exhibits the linear Regge trajectory
at the leading order in $1/\lambda$ expansions, with a non-linear term
added as a subleading correction.
By analyzing the discrete symmetries of our brane configuration,
we read off the parity and charge conjugation quantum number of
the open string states.
It looks quite non-trivial
that many of the open string states found in our analysis can be
identified with the mesons found in the experiments. 
This result can be regarded as an evidence of the validity
of the gauge/string duality including the massive excited string
states. Thorough investigations toward this direction would be
worthwhile.
For example, it would be nice to work out the interactions
that involve such excited meson from string theory.
Similar analysis in the closed string sector would also
be interesting.

There remain several important problems to clarify. 
{}First, as mentioned in \S \ref{review} and \S \ref{data}, 
the value of $\alpha'$ is not in good agreement with the expected value
if we use (\ref{alpha}) and (\ref{mkk}). It would be interesting
to see if it is improved by taking into account possible
corrections.
It is also important to calculate corrections
to our mass formula (\ref{Mn1}) to see how much our results are affected
by them.
In particular, since most of the mesons we have
been discussing are of the same order or heavier than the mass
scale $\sim\mkk$ of the artifacts of the model, one might
think that the deviation from realistic QCD would be significant.
However, we know from many other nice results in holographic QCD
that the effect of these artifacts seems to be much smaller than what
one would naively expect. It is probable that our results described in \S \ref{data}
will not be changed much by taking into account the corrections.
The situation may be analogous to the fact that the quenched
approximation works very well in lattice QCD.
Anyway, computation of the corrections should be done to justify all these.

Another related problem is the existence of the open string states that
cannot be identified with the mesons in QCD.
In \S \ref{SO5}, we restricted our attention to the states that are
invariant under the $SO(5)$ and $\bZ_2$ symmetry to
exclude such artifacts from our consideration as many as we can.
However, one should be aware that
although this is a necessary condition to obtain mesons in QCD,
it may not be a sufficient condition, and some of the states obtained
in this paper might contain the artifacts of the model.
To completely get rid of all the artifacts, we
need to take a limit analogous to the continuum limit in lattice QCD,
as discussed in \S \ref{review}.
To this end, we have to extrapolate our analysis to small $\lambda$
region, in which supergravity approximation breaks down.

As pointed out in \S \ref{data}, our model predicts the states that
have no candidate in Table \ref{table1}.
Among them are the second excited 
massive modes with $J^{PC}=0^{++}, \,1^{+-}$, and $2^{--}$, whose masses are
expected to be approximately 1700 MeV.
This fact does not lead to the conclusion that 
our model contradicts with the results of experiments immediately,
since these states may correspond to wide resonances
that are difficult to observe.
If so, it would be interesting to compare our results with
the predictions from the quark model.
(For a review, see the section of ``Quark model'' in Ref.~\citen{PDG}.)
In fact, it is known that the mesons with
 $J^{PC}=0^{++}$, $1^{+-}$, and $2^{--}$ can arise from
the quark model.
The lightest $q\ol q$ states with $J^{PC}=0^{++}$ and $1^{+-}$
are those labeled as $\hat n^{2S+1}L_J=1^3P_0$ and $1^1P_1$,
and are identified with $a_0(1450)$ and $b_1(1235)$, respectively.
Here, $S$, $L$, $J$, and $\hat n$ are the total spin carried by the
constituent quark and anti-quark, orbital angular momentum, spin of the
meson, and the principal quantum number corresponding to the
radial excitation, respectively.
Then, it is natural to identify
the second lightest $J^{PC}=0^{++},1^{+-}$ states
and the lightest $J^{PC}=2^{--}$ state in the quark model
with the $J^{PC}=0^{++}$, $1^{+-}$, and $2^{--}$ states
found as the open string states with $(N,n)=(2,0)$ in (\ref{N=2}).
They are labeled as
$\hat n^{2S+1}L_J=2^3P_0$, $2^1P_1$, and $1^3D_2$,
respectively. The masses of these states estimated in the quark model
are indeed all close to 1700 MeV. (See, for example, Refs.~\citen{GoIs}
and \citen{EbFaGa}.)

\section*{Acknowledgements}

We would like to thank our colleagues at the Institute
for the Physics and Mathematics of the Universe (IPMU)
and the particle theory group at Nagoya University
for helpful discussions. S.S. is especially grateful to
Z. Bajnok, K. Hori, R. A. Janik, and T. Onogi for valuable discussions.
T.S. thanks H.~Fukaya, M.~Harada, and M.~Tanabashi for valuable discussions.
The work of S.S. is supported  in part by 
a Grant-in-Aid for Young Scientists (B), the Ministry of Education, 
Culture, Sports, Science and Technology (MEXT), Japan,
a JSPS Grant-in-Aid for Creative Scientific Research
(No. 19GS0219), and also by World Premier International
Research Center Initiative (WPI Initiative), MEXT, Japan. 
We are also grateful to the organizers of the workshop ``Summer
Institute 2008'', where part of this work was done.

\appendix

\section{Higher Excited States}
\label{higher}

The second excited states with $N=2$ belong to
$\fnd_{\,9}\otimes \Tasym_{\,84}\oplus
\fnd_{\,9}\otimes  \sym_{\,44}$ of
the little group $SO(9)_{1\sim 9}$. (See section 5.3 of Ref.~\citen{GSW})
The fields corresponding to the particles in
 $\fnd_{\,9}\otimes\Tasym_{\,84}$ and 
$\fnd_{\,9}\otimes  \sym_{\,44}$ are denoted as
$B_{A_1[A_2A_3A_4]}$ and $C_{A_1(A_2A_3)}$
with ($A=1,\dots,9$), respectively. Here, $[A_1,\dots,A_n]$ and
$(A_1,\dots,A_n)$ denote anti-symmetrization and symmetrization
of the indices $A_1,\dots,A_n$, respectively.

The $SO(4)_{6\sim 9}$
invariant states are
\begin{eqnarray}
B_{\alpha_1[\alpha_2\alpha_3\alpha_4]}\ ,~~
B_{a[a\alpha\beta]}\ ,~~
\epsilon^{a_1a_2a_3a_4}
B_{a_1[a_2a_3a_4]}\ ,~~
C_{\alpha_1(\alpha_2\alpha_3)}\ ,~~
C_{a(a\alpha)}\ ,~~C_{\alpha(aa)}\ ,
\end{eqnarray}
where $\alpha,\beta=1,\dots,5$ and $a=6,\dots,9$.
The $\tau$-parity invariant components are
\begin{eqnarray}
&&B_{M_1[M_2M_3M_4]}\ ,~~
B_{MN}^{[1]}\equiv B_{a[aMN]}\ ,~~
B_{MN}^{[2]}\equiv B_{y[yMN]}\ ,~~
C_{M_1(M_2M_3)}\ ,\nn\\
&&C_M^{[1]}\equiv C_{y(yM)}\ ,~~
C_M^{[2]}\equiv C_{M(yy)}\ ,~~
C_M^{[3]}\equiv C_{a(aM)}\ ,~~
C_M^{[4]}\equiv C_{M(aa)}\ ,
\end{eqnarray}
where $M=1,\dots,4$.
The lightest meson fields obtained from these fields together with
their $J^{PC}$ are listed as follows:
\begin{align}
&B^{(0)}_{j[i_1i_2i_3]}~(1^{+-})\ ,~~
B^{(0)}_{z[i_1i_2i_3]}~(0^{++})\ ,~~
B^{(0)}_{i[i_1i_2z]}~(2^{++},1^{++},0^{++})\ ,~~
B^{(0)}_{z[i_1i_2z]}~(1^{+-})\ ,\nn\\
&B^{[1,2](0)}_{ij}~(1^{+-})\ ,~~
B^{[1,2](0)}_{iz}~(1^{++})\ ,~~
C^{(0)}_{i_1(i_2i_3)}~(3^{--},2^{--},1^{--})\ ,\nn\\
&C^{(0)}_{z(ij)}~(2^{-+})\ ,~~
C^{(0)}_{i(zj)}~(2^{-+},1^{-+},0^{-+})\ ,~~
C^{(0)}_{i(zz)}~(1^{--})\ ,\nn\\
&C^{(0)}_{z(zi)}~(1^{--})\ ,~~C^{(0)}_{z(zz)}~(0^{-+})\ ,~~
C_i^{[1\sim 4](0)}~(1^{--})\ ,~~C_z^{[1\sim 4](0)}~(0^{-+})\ ,
\label{N=2}
\end{align}
where $i,j=1,2,3$.
Therefore, we have $3^{--}$, $2^{++}$,
two $2^{-+}$, $2^{--}$, seven $1^{--}$,
four $1^{+-}$, three $1^{++}$, $1^{-+}$, 
two $0^{++}$, six $0^{-+}$, as the lightest modes.

The excited states with $N\ge 3$ can be analyzed in a
similar way.
The classification of the excited states with respect to the
$SO(9)$ little group can be obtained using the techniques
used in Refs.~\citen{Bianchi:2003wx} and \citen{Beisert:2003te}.

\section{Discrete Symmetry}
\label{Z2}

In this Appendix, we classify the $\bZ_2$ symmetries of
the Witten's D4-brane background with probe D8-branes. 
Type IIA string theory has $\bZ_2$ symmetries 
generated by $I_{\rm even}$, $I_{\rm odd}\Omega$, and $(-1)^{F_L}$,
where $I_{\rm even}$ and $I_{\rm odd}$ are
spatial involutions that flip the sign of
even and odd numbers of coordinates,
respectively, $\Omega$ is the world-sheet parity transformation,
and $F_L$ is the left moving space-time fermion number.\footnote{
We do not care much about $(-1)^{F_s}$, where $F_s$ is the space-time
fermion number, since its action is trivial on the mesons.
}
$(-1)^{F_L}$ acts on the fields in R-R sector and R-NS sector
as multiplication by $-1$, and hence,
it does not keep the RR 4-form field strength of 
the background (\ref{RR}) invariant.
Instead, $P_\tau\equiv I_{y9}(-1)^{F_L}$
is a symmetry of the background, as it keeps the $F_4$ flux
as well as the metric invariant.
Here, $I_{i_1i_2\dots i_n}$ denotes the involution
$(x^{i_1},x^{i_2},\dots,x^{i_n})\ra
(-x^{i_1},-x^{i_2},\dots,-x^{i_n})$.
In \S \ref{CP}, we have learned two other $\bZ_2$ symmetries
corresponding to parity and charge conjugation
symmetries generated by $P\equiv I_{123z}$ and
$C\equiv I_{z89}\Omega(-1)^{F_L}$, respectively.
The spatial involutions  $I_{\rm even}$
that generate $\bZ_2$ subgroups of
$SO(1,3)\times SO(5)$ isometry are of course symmetries of the system.
On the other hand, the spatial involutions $I_{\rm even}$
that involve $y\ra -y$ are not symmetries of the system, since they map
the D8-branes placed at $y=0$ to \AD8-branes.
However, if we combine it with $(-1)^{F_L}$, which
maps \AD8-branes back to D8-branes, the D8-branes are kept
invariant. Therefore, $P_\tau$ is a symmetry of the system.
It is then easy to show that the $\bZ_2$ symmetries of the system we
should consider are those generated by the combinations of $P_\tau$,
$P$, $C$, and elements of a $\bZ_2$ subgroup of
$SO(1,3)\times SO(5)$ isometry.

$P_\tau$ defined above acts in the same way as ``$\tau$-parity''
considered in Ref.~\citen{BrMaTa}. Let us show that the
quarks and gluons are invariant under this $\tau$-parity.
To this end, we turn to the D4/D8/\AD8-brane configuration.
As briefly reviewed in \S \ref{review},
QCD is realized on the D4-brane world-volume in the D4/D8/\AD8-brane
system embedded in a flat ten-dimensional space-time
$\bR^{1,3}\times S^1\times\bR^5$.
The coordinates of the $\bR^{1,3}$ and $S^1$ factors correspond to
the coordinates $x^\mu$ and $\theta$ used in the background
(\ref{metric}), respectively, and
the radial and angular directions of the $\bR^5$ factor
correspond to $r$ and $S^4$, respectively.
In this Appendix, we parametrize the $\bR^5$ factor by
 $(x^5,\dots,x^9)$.
Although these coordinates are not the same as those used in
\S \ref{SO5}, the involution $x^9\ra -x^9$
acts in the same way.
In this picture,
the D4-branes are extended along $\bR^{1,3}$ and $S^1$ directions
and placed at the origin of $\bR^5$,
while the D8-branes and \AD8-branes are extended along
$\bR^{1,3}\times\bR^5$ directions and placed at $\theta=\pi/2$
and $\theta=-\pi/2$, respectively.
The gluons are created by 4-4 strings
and the quarks are created by 4-8 and 4-$\ol 8$ strings,
where a $p$-$p'$ string is an open string stretched between
D$p$-brane and D$p'$-brane.

The involution $y\ra -y$ corresponds to the involution acting
on the $S^1$ as $\tau\ra-\tau$, where
we have defined $\tau\equiv\theta-\pi/2$. Therefore,
$P_\tau$ acts as $I_{\tau 9}(-1)^{F_L}$
in this D4/D8/\AD8 system.
To know how $P_\tau$ acts on
gluons and quarks, it is convenient to
T-dualize the system along the $S^1$ direction and
consider a D3/D9/\AD9 system.\footnote{
Because we impose the anti-periodic boundary condition
for the fermions along the $S^1$, the T-duality along this
$S^1$ yields Type 0B string theory.
This boundary condition is not essential in our argument,
since we are only interested in the transformation property
of the open strings under $P_\tau$.
} 
By parametrizing the T-dualized $S^1$ by $\wt\tau$,
the $\tau$-parity $P_\tau$ can be interpreted as
a 180-degree rotation in the $\wt\tau$-$x^9$ plane
around the D3-brane.
Since the gauge field on the D3-brane and the massless
fermions in the spectrum of the 3-9 and 3-$\ol 9$ strings
are invariant under the rotation in the $\wt\tau$-$x^9$ plane,
we conclude that the gluons and quarks are invariant
under the $\tau$-parity $P_\tau$.

\section{Contribution of RR Field to the Mass Formula}
\label{App:RR}

Since the radius of the $S^4$ is proportional to $\mkk^{-1}$,
the background RR 4-form field strength $F_4$ (\ref{RR}) 
in our convention is proportional
to $N_c \mkk^4$.\footnote{See also Appendix A of Ref.~\citen{SS1} for our
convention of the RR fields.} Then, a possible mass term 
for open string modes $\Phi$ and $\Phi'$ induced by the RR background
on the D8-brane world-volume will be schematically written as
\begin{eqnarray}
 \int d^9x\,\alpha'g_s F_4\,{\rm Tr}(\Phi\, \Phi')
\sim  \int d^9x\,\lambda^{1/2}\mkk^2{\rm Tr}(\Phi\, \Phi')\ .
\label{massmat}
\end{eqnarray}
Here, we assume that the kinetic terms of $\Phi$ and $\Phi'$ are
canonically normalized, and the Lorentz indices of $F_4$, $\Phi$, and
$\Phi'$ are properly contracted. We only consider the isovector mesons,
for which there are no mixing terms with closed string states.
In our convention, the $g_s$ dependence of the effective action
agrees with the string loop expansion if it is written in terms of
$F_4'=g_s F_4$. Since the dimensionless combination $\alpha'^2 F'_4$
is of $\cO(\lambda^{-1/2})$, the higher order terms with respect to
the background RR 4-form field strength can be neglected.
The mass matrix element obtained from (\ref{massmat})
is of $\cO(\lambda^{1/2})$, which is of the same order as 
$\cO(\lambda^{1/2})$ term in (\ref{Mn1}). Thus, one might think
that the $\cO(\lambda^{1/2})$ term in our mass formula (\ref{Mn1})
would be modified by taking into account the effect of the
background RR field.

However, it can be shown that the terms like (\ref{massmat})
can only be possible when $\Phi$ and $\Phi'$ are not in the same
excitation level. Then, by diagonalizing the mass matrix,
it is easy to see that the contribution of the mixing term (\ref{massmat})
to the mass squared of mass eigenstates is at most of $\cO(1)$.
Therefore, the contribution of $F_4$ in the background in
the equation of motion (\ref{eom}) can be neglected.

To see that $\Phi$ and $\Phi'$ cannot be in the same excitation level,
it is convenient to T-dualize the system along the $y$ direction
as we did in \S \ref{SO5}. Then, the D8-branes are mapped
to D9-branes and the RR 4-form field strength $F_4$ is mapped
to an RR 5-form field strength $F_5$. We use the same notation
$\Phi$ and $\Phi'$ for the open string modes on the D9-brane.
Recall that type IIB string theory
with D9-branes is invariant under the world-sheet parity transformation
$\Omega$. Then, because $F_5$ is odd under $\Omega$,
${\rm Tr}(\Phi\Phi')$ should also be odd under $\Omega$
for allowing an interaction term as $F_5 {\rm Tr}(\Phi\Phi')$.
This implies $(-1)^{N+N'}=-1$, where
$N$ and $N'$ are the excitation levels of $\Phi$ and $\Phi'$,
respectively, and hence, $\Phi$ and $\Phi'$ cannot be in the
same excitation level.

\end{document}